# Novel Quantum Circuit Designs of Random Injection and Payoff Computation for Financial Risk Assessment

*Yu-Ting Kao, Yeong-Jar Chang, Ying-Wei Tseng*
*Industrial Technology Research Institute*

**ABSTRACT**

Quantum entanglement enables exponential computational states, while superposition provides inherent parallelism. Consequently, quantum circuits are theoretically capable of supporting large-scale parallel computation. However, applying them to financial analysis—particularly in the areas of random number generation and payoff computation—remains a significant challenge. Experts generally believe that quantum computing relies on matrix operations, which are deterministic in nature without randomness. This inherent determinism makes it particularly challenging to design quantum circuits that require random number injection. JP Morgan [1] introduced the piecewise linear (PWL) approach for modeling payoff computations but did not disclose a quantum circuit capable of identifying values exceeding the strike price, suggesting a possible reliance on classical pre-processing for interval classification.

This paper presents an integrated quantum circuit with two key components: one for random number injection, applicable to risk assessment, and the other for direct payoff computation, relevant to financial pricing. These components are compatible with a scalable framework that leverages large-scale parallelism and Quantum Amplitude Estimation (QAE) to achieve quadratic speedup. The circuit was implemented on IBM Qiskit and evaluated using 8 parallel threads and 1,600 measurement shots. Results confirmed both the presence of randomness and the correctness of payoff computation. While the current implementation uses $2^3$ threads, the design is scalable to $2^n$ threads for arbitrarily large $n$, offering a potential path toward demonstrating quantum supremacy.

## 1. INTRODUCTION

JP Morgan [1] proposed a quantum circuit for mean computation, though loading arbitrary probability density function (PDF) into quantum computers remains a challenging problem. It also introduced the PWL approach for modeling payoff computations. Building on this foundation, Wilkens [2] introduced the use of classical preprocessing to partition data into PWL regions for subsequent processing, particularly suitable for risk measurement, while Vazquez [3] proposed analog parallel parameter multiplication and addition circuits that significantly reduce the circuit complexity than digital parallel designs. More recently, SNUG Taiwan [4] developed mixed-signal quantum circuits for option pricing, combining the simplicity of analog design with the flexibility and synthesizability of digital circuits. This approach significantly reduces circuit complexity, enhances noise resistance, and enables large-scale simulations of stock price volatility. The study demonstrated that large-scale parallelism can be achieved not only with digital circuits but also with analog designs.

However, current payoff computations still depend on classical preprocessing to isolate regions above the strike price before quantum evaluation. A fully quantum architecture capable of directly handling the entire payoff computation remains undeveloped. Addressing this gap is the first focus of this work.

Beyond this, randomness plays a critical role in financial analysis, particularly in simulating uncertain events like asset price fluctuations and risk exposure. Quantinuum [5] demonstrated that quantum systems can generate provably random numbers, supporting applications such as Woerner and Egger[6] QAE-based risk analysis, which outperforms classical Monte Carlo simulations in computing Value at Risk (VaR) and Conditional Value at Risk (CVaR). Building on this, Matsakos and Nield [7] introduced a quantum Monte Carlo framework showing that quantum superposition can model stochastic behavior, while Montanaro [8] explored quantum acceleration of Monte Carlo methods, achieving near-quadratic speedups for expectation estimation.

Despite these advances, a core dilemma remains: quantum operations are deterministic prior to measurement. Although QAE offers a quadratic speedup, it suppresses randomness. Conversely, recovering statistical randomness through repeated measurements negates the quantum advantage. Designing circuits that enable both randomness and quantum speedup is therefore difficult. The financial risk assessment often requires processing millions of parallel threads, each influenced by millions of random factors. Achieving both large-scale parallelism and inherent randomness is more challenging for scalable quantum financial analysis.

This paper proposes two novel quantum circuit components: one for random number injection and another for payoff computation. These components can be integrated into a unified framework that leverages large-scale parallelism and QAE to achieve both quantum speedup and potential quantum supremacy. For example, the baseline quantum circuit can process one million threads in parallel. With the integration of our design, one million random factors can simultaneously influence these thread-level computations. The second component further enables efficient payoff evaluation across all threads. Finally, QAE requires only 32 circuit operations (denoted by $m$) to achieve the same level of accuracy as 1,024 times (denoted by $m^2$) of classical Monte Carlo simulations, thus demonstrating the quantum quadratic advantage. The large number of one million threads highlights the potential for quantum supremacy.

In this paper, we shall introduce more details on how random injection can achieve quadratic speedup and quantum supremacy. This revolutionary method extends its applicability to numerous domains requiring extensive random simulations, including financial risk assessment, where complex market behaviors must be modeled; physical system modeling, where stochastic processes govern system dynamics; stochastic optimization in machine learning, where uncertain parameters affect model performance; and cryptographic analysis, where random number generation and statistical properties are critical for security protocols.

## 2. METHODOLOGY

### 2.1. Random Injection

**Figure 1** illustrates the original 8 threads operating in massive parallel processing. After incorporating random numbers, the system maintains 8 threads in massive parallel processing while using random number generation to create 4 types of combinations, $r_0r_1=(00, 01, 10, 11)$, such that $\frac{1}{4}$ probability (11) will cause E to change sign (e.g., 0→1), while $\frac{3}{4}$ probability (00, 01, 10) remains unchanged.

**Figure 2** demonstrates that random numbers simultaneously affect all 8 threads ($ABC = 2^3$), transforming the original 8 outputs into 16 outputs. From the output results, the probabilities are altered to $\frac{1}{4}$ and $\frac{3}{4}$, clearly showing the "random" effect on parallel computation for risk asset evaluation.

In order to investigate the equivalence between Random Injection and Grid Sampling methods in expectation value computation using quantum computing, we demonstrate that the measurement results by QAE can effectively represent the statistical outcomes of large-scale random simulations. For example, for a system x with 266 qubits, we need to compute the population mean of g(x) among all $2^{266}$ threads. Classical computers cannot handle such enormous computational demands and has to rely on random simulation methods (e.g., Monte Carlo with 10,000 samples) to approximate the population mean.

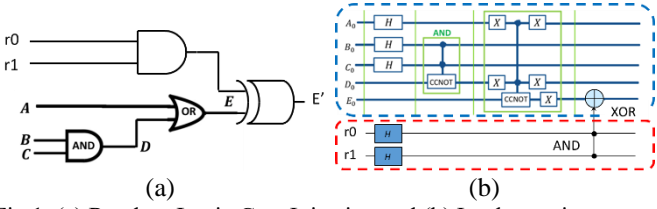

Fig 1. (a) Random Logic Gate Injection and (b) Implementing Random Injection with Qiskit

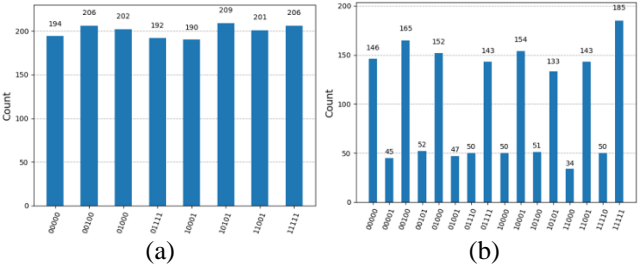

Fig 2. (a) Original circuit: 8 outputs and (b) Original circuit after adding Random injection: 16 outputs

**Method 1: Random Injection** defines the observation function $g(x) = f(x)\, \text{XOR}\, (r == 3)$ where r is a random variable following a discrete uniform distribution $r \in \{0, 1, 2, 3\}$, with each value having probability $\frac{1}{4}$. The execution flow involves input data of $2^{266}$ threads, injecting random number r during computation, and computing average after completion. The expectation value calculation shows:

$$E[g(x)] = E[f(x)\, XOR\, (r == 3)]$$
$$= \frac{1}{4} \times [f(x)\, XOR\, 0 + f(x)\, XOR\, 0 + f(x)\, XOR\, 0 + f(x)\, XOR\, 1]$$
$$= \frac{3}{4} \times f(x) + \frac{1}{4} \times (f(x)\, XOR\, 1)$$

**Method 2: Grid Sampling** defines the observation function $g(x, r) = f(x)\, \text{XOR}\, (r = 3)$ where r is a deterministic grid parameter, $\text{Grid} = r \in \{0, 1, 2, 3\}$, expanding the total thread count to $2^{268}$. The execution flow involves input data with grid parameters forming $2^{268}$ threads, executing computation without randomness, and computing average after completion. The exhaustive calculation shows:

$$E[g(x,r)] = \frac{1}{4} \times \Sigma[r = 0\, to\, 3]\, g(x,r)$$
$$= \frac{1}{4} \times [g(x,0) + g(x,1) + g(x,2) + g(x,3)]$$
$$= \frac{1}{4} \times [f(x) + f(x) + f(x) + (f(x)\, XOR\, 1)]$$
$$= \frac{3}{4} \times f(x) + \frac{1}{4} \times (f(x)\, XOR\, 1)$$

The equivalence theorem states that the expected value of Random Injection and the exhaustive value of Grid Sampling methods are identical:
$$E[g(x)] = E[g(x,r)] = \tfrac{3}{4} \times f(x) + \tfrac{1}{4} \times (f(x)\, XOR\, 1)$$

QAE has three key features that illustrate how it can solve complex statistical problems. First, QAE is deterministic, unlike traditional random sampling methods, which involve inherent variability. Second, QAE ensures that the measurement result directly corresponds to the exact average over all possible outcomes, as would be obtained through exhaustive evaluation over deterministic grid points, without requiring repeated sampling or statistical approximation. Third, the result produced by QAE is also equivalent to the expected value from computations involving large-scale random injections, as it is mathematically identical to the deterministic average.

The term "random injection" in massive parallelism is often mistakenly assumed to follow the same convergence rate as traditional Monte Carlo methods, i.e., $O(m^{-1/2})$ [6]. However, our proposed approach integrates random injection with QAE measurements, resulting in an improved convergence rate of $O(m^{-1})$, where $m$ denotes the number of quantum operations. This enhancement arises from the aggregation of all quantum states to compute the expected value of randomized massive functions, rather than relying on classical sampling-based iterations for direct output measurements.

This study demonstrates that QAE enables accurate estimation of large-scale random simulations, making quantum computation a deterministic statistical estimator. The demonstrated statistical reliability of QAE highlights its potential as an efficient alternative to conventional financial risk analysis in large-scale quantum simulations across diverse application domains.

### 2.2. Payoff computation

The design aims to create a quantum circuit (**Fig. 3**) that processes PDF values ranging from 0 to 31, automatically ignoring the values smaller than 24 and calculating the average of values between 24 and 31. These PDFs can come from direct inputs or large-scale parallel computations. Traditionally, the PWL method uses classical computers to filter results greater than or equal to 24 and then sends the data to quantum computers for averaging, that may suffer from classical bottlenecks and communication delays. To address this, our approach uses **Mixed-Signal Switch** technology to perform both filtering and averaging within the purely quantum framework. For price encoding, such as a strike price of 900 represented as 11000 in binary, the Mixed-Signal Switch recognizes that the first few bits must be 1 to meet the threshold. This approach calculates the weighted average of strike prices between 900 and 970, corresponding to the binary range from 11000 to 11111, thus completing the payoff computation for the strike price 900, represented by 11000(binary) or 24(decimal). Unlike the JP Morgan circuit, which averages all 32 threads, our method uses AB=11 as a digital control switch to route the LM qubit to another PF qubit if the price ≥ 24 threshold, enabling selective quantum processing for the payoff computation. The integration of digital control and analog quantum operations through the Mixed-Signal Switch can identify whether the price exceeds the strike price, enabling real-time payoff calculations within a purely quantum circuit.

The circuit is designed with a fixed strike price of 24. However, if other strike prices are needed, as shown in **Figure 3**, all input prices can be shifted accordingly to generate the desired strike price. For example, to obtain a strike price of 26, we can pre-shift all inputs by subtracting 2 to each *x*, since the desired expression *x*−26 is equivalent to (*x*−2)−24, which is what the circuit computes. This demonstrates that the proposed circuit constitutes a scalable architecture that can be adapted to different financial scenarios and strike price requirements through simple addition or subtraction operations, without requiring circuit redesign.

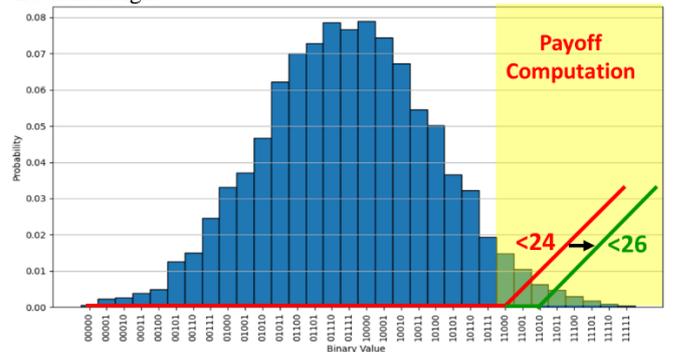

Fig 3. The quantum circuit automatically filters and averages the PDF

**Figure 4** presents a significant advancement in random injection and payoff computation for financial analysis. The other components of the circuit, such as multi-thread operations and weighted averaging, are well-established. First, it enables massive parallel randomness without losing quantum advantages. Second, it overcomes the limitation that quantum payoff could only be performed after classical pre-processing on PWL classifications.

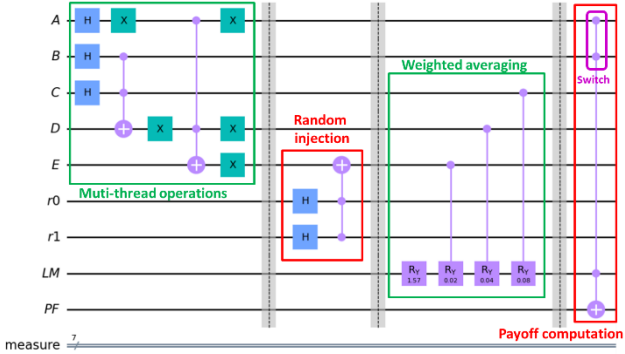

Fig 4. Quantum Circuit Architecture for Financial Analysis

## 3. EXPERIMENTAL RESULTS

### 3.1. Random Injection Histogram Analysis Verification

It compares the quantum circuit results with the expected E sign-change probability of 0.25. Under the condition of 1600 shots, the average error is 9.26%(**Fig.2**), with the error attributed to randomness effects.

This research confirms that a single circuit can generate $2^n$ groups (n=3 in this case) of massive parallel computation, with each group producing different random numbers. Based on this research, the 8-threads system can be extended to $2^{266}$ threads.

### 3.2 Average calculation result

This proposed circuit can input all stock price calculation results, and automatically filter the specified PDF in the quantum system through the switch mechanism to complete the payoff calculation. In this experiment, through Monte Carlo simulation with 1600 iterations, we use 5-bits (Price=ABCDE) to represent stock price variations. The histogram shows different stock prices (Price=0, 1, 2, ... 31) and their corresponding occurrence frequencies. With all qubits A~E, there are 32 possible stock price results, of which 16 stock prices will appear while the other 16 stock prices have zero probability of occurrence. The x-axis represents gradually increasing stock prices, and the y-axis represents occurrence frequencies. To obtain the average stock price over 1600 iterations, a weighted average calculation is required.

Quantum computation results are easily converted to probabilities for comparison, where probability p is the square of parameter Ci, expressed as $p=|C_i|^2$. In **Figure 4**, Payoff(PF) and Low-Bit Mean(LM) are calculated based on the weighted average calculation method of JP Morgan [1], as follows:

$$PF = \sum_{i=0}^{7} P[i+24] \cdot (i \cdot m + k) \quad (1)$$
$$LM = \sum_{i=0}^{7} (P_i + P_{i+8} + P_{i+16} + P_{i+24}) \cdot (i \cdot m + k) \quad (2)$$

where $P_i$ are the probabilities and m, k are constant

The original computation involves 8 threads, which expand to 16 after random injection. These then undergo payoff computation through the proposed quantum circuit. The switching mechanism functions as follows: PF is set to LM if AB = 11 (corresponding to prices ≥ 24); otherwise, PF is set to 0. Experimental results in **Figure 5** demonstrate that the circuit computes the weighted averaging only when the stock price exceeds the threshold of 24, assigning PF = 0 in all other cases. This design eliminates the need for classical preprocessing and enables real-time conditional payoff evaluation based on price thresholds, resulting in a more efficient and streamlined quantum financial computing process.

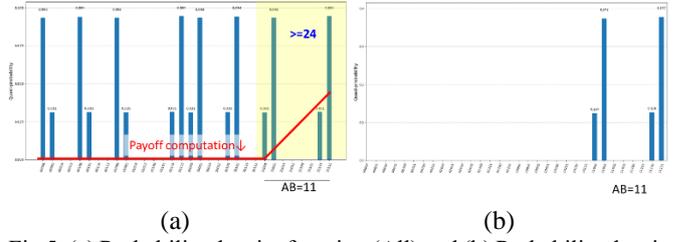

Fig 5. (a) Probability density function (All) and (b) Probability density function after payoff(PF)

The original idea was to align the desired value $\sqrt{x}$ with the computed result $\sin(y)$ from the quantum circuit, under the condition that $x = y$, and the values lie within a small neighborhood around $\frac{\pi}{4}$. This observation suggests a potential angular relationship between the square root and the sine function. Thus, we hypothesize a scaling relationship involving a factor of 2 in angular variation, expressed as:

$$x = 0.785 + m \times i \times \theta, \quad y = 0.785 + i \times \theta, \quad \text{with } m=2$$

However, a prior study [4] demonstrated that the minimal error occurs when the scaling factor $m=1.57$, leading to the refined expression: $x = 0.785 + 1.57 \times i \times \theta$, where $\theta$ is the rotation angle in quantum circuit. Applying a linear transformation by dividing only the right-hand side by 1.57 yields: $x = 1/2 + i \times \theta$. This result exactly matches the expression derived from the Taylor series in [1], thereby providing theoretical consistency across different formulations.

These experimental results present a comparative analysis using four bar chart groups for $\theta = 0.01$ (**Fig.6**):
(1) Classical computation with input $x=0.785+0.0157i$, serving as the baseline.
(2) Original quantum computation using $y=x=0.785+0.0157i$.
(3) Quantum computation with analog calibration, using $y=0.785+0.01i$ based on a scaling factor $m=1.57$.
(4) Quantum computation based on the Taylor series approximation, also using $y=0.785+0.01i$.

The corresponding classical computation in case (4) uses the input $x=1/2+0.01i$, which is equivalent to the baseline divided by 1.57. Consequently, cases (3) and (4) yield the identical results with similarly small error margins. These findings support two key observations:

First, the analog calibration method proposed in [4] and the Taylor series-based approach in [1] both reduce the approximation error of PF (payoff) from 3.912% to 0.0134%, demonstrating their mathematical equivalence in this context. The LM (lower-bit mean) results also verify the consistency between quantum and classical computations, showing an error of 3.248%, reduced to 0.0079% after analog calibration.

Second, the proposed quantum circuit is capable of executing payoff computations entirely within the quantum circuit, without requiring classical pre-analysis or piecewise linear (PWL) segmentation. This highlights the circuit's efficiency and integration potential for fully quantum financial simulations.

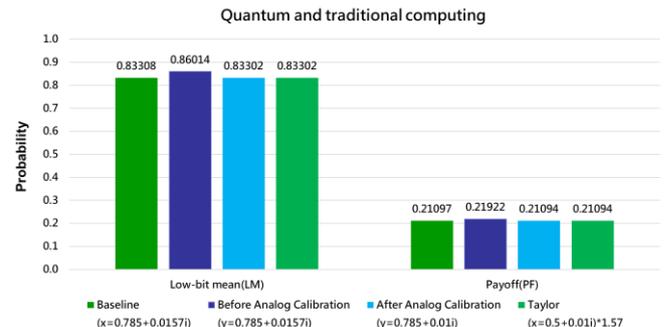

Fig 6. Comparison of Quantum and traditional computing results

## 4. CONCLUSION

This paper presents two innovative quantum circuits, Random injection and Payoff computation, specifically designed for financial risk simulation applications. By utilizing the deterministic computation capabilities of quantum circuits through Quantum Amplitude Estimation (QAE), we successfully approximate stochastic operations. Our invention combines the benefits of extensive randomness with quantum quadratic speedup.

Empirical validation of our approach shows remarkable computational accuracy. For instance, quantum circuit results obtained after random injection with 1,600 shots demonstrate an average error of only 9.26% when compared to the expected PDF. This proves the reliability of quantum random injection for financial risk assessment. Furthermore, quantum computation can independently complete Payoff computations without requiring any traditional computational preprocessing, ensuring complete consistency with classical computation results and eliminating the hybrid computational dependencies that have historically limited the application of quantum methods in finance.

A comparative analysis of average payoff results from PDF inputs reveals the precision improvements achieved through our methodology. The original Payoff error of 3.912% was dramatically reduced to 0.0134% after calibration.

The mathematical validation demonstrates that the Taylor series analysis equals the calibrated result, validating the effectiveness of the 1.57 multiplication factor in maintaining computational precision while successfully bridging the gap between quantum and classical computational paradigms.

Our work effectively integrates random injection and payoff computation with quantum speedup achieved through QAE. Both components support large-scale quantum parallelism, contributing to the potential realization of quantum supremacy. This integration lays a robust and practical foundation for the development of quantum financial risk assessment systems capable of addressing complex real-world financial analyses. It enhances the practical utility of quantum computing in financial technology applications and opens new opportunities for quantum-native financial modeling, overcoming the limitations of classical computation in both speed and scalability.